\documentstyle[aps,preprint,prb]{revtex}
\begin{document}
\draft
\preprint{MIC-DFM preprint}
\title{Electron--electron
scattering in linear transport in two-dimensional systems}

\author{Ben Yu-Kuang Hu}

\address{Mikroelektronik Centret,
Bygning 345 \o st, Danmarks Tekniske Universitet,
DK-2800 Lyngby, Denmark}

\author{Karsten Flensberg}

\address{Mikroelektronik Centret,
Danmarks Tekniske Universitet,
DK-2800 Lyngby, Denmark and
Dansk Institut for Fundamental Metrologi,
Bygning 307, Anker Engelunds Vej 1,
DK-2800 Lyngby, Denmark}

\date{\today}
\maketitle

%%%%%%%%%%%%%%%% ABSTRACT %%%%%%%%%%%%%%%%%%%%%%%%%%%%%%
\begin{abstract}

We describe a method for numerically incorporating electron--electron
scattering in quantum wells for small deviations of the distribution
function from equilibrium, within the framework of the Boltzmann equation.
For a given temperature $T$ and density $n$, a symmetric matrix needs
to be evaluated only once, and henceforth it can be used to describe
electron--electron scattering in any Boltzmann equation linear-response
calculation for that particular $T$ and $n$.
Using this method, we calculate the
distribution function and mobility for electrons in a quantum-well,
including full finite-temperature dynamic screening effects.
We find that at some
parameters which we investigated, electron--electron scattering
reduces mobility by approximately 40\%.

\end{abstract}


%%%%%%%%%%%%%%%% INTRODUCTION %%%%%%%%%%%%%%%%%%%%%%%%%%%%%%
\narrowtext
\section{Introduction}
\label{sec:intro}

The effect of electron--electron ($e-e$) interactions on transport
in bulk-like systems of various dimensionalities is still an area of
active research to this day, both experimentally\cite{mole94,chab95} and
theoretically\cite{sanb95,schu95}.  On the face of it, it would seem
that in the case of doped parabolic band semiconductors
where umklapp processes are negligible, $e-e$ scattering should not
affect the linear transport properties of bulk-like systems
(purely quantum effects such as weak-localization corrections
excepted) since an $e-e$ scattering event conserves the total current
in the system.  Nevertheless, it has been appreciated for a long time
that $e-e$ scattering {\em can} affect
the mobility of a system semiclassically
by scattering carriers into or out of parts of
the Brillouin zone which are strongly affected by the other
available scattering mechanisms.\cite{herr56,keye58}

In Ref. \onlinecite{herr56}, it was shown that the expression for mobility
in the presence of quasi-elastic scatters takes on different forms
in the limits of zero and infinitely strong $e-e$ scattering.
Given a quasi-elastic energy-dependent transport
(i.e., calculated with the $1-\cos\theta$ term)
scattering time $\tau(\varepsilon)$ due to other scattering processes
in the system such as acoustic phonons or impurities, the
mobilities of the system for the cases of zero and infinite
$e-e$ scattering rates, respectively,
are given by
\begin{eqnarray}
\mu_0 &=& \frac{e\langle\tau\rangle}{m};\nonumber\\
\mu_\infty &=& \frac{e}{\langle \tau^{-1} \rangle m}.
\end{eqnarray}
Here,
\begin{equation}
\langle A\rangle = {2\over n}\int \frac{d{\mbox{\boldmath{$k$}}}}{(2\pi)^d}\
\varepsilon(\mbox{\boldmath{$k$}}) \left( -\frac{\partial f_0}{\partial
\varepsilon}
\right) A(\mbox{\boldmath{$k$}}),
\end{equation}
where $n$ is the carrier density of the system,
$\varepsilon(\mbox{\boldmath{$k$}})
= \hbar^2 k^2 /(2m)$, $f_0(\varepsilon(\mbox{\boldmath{$k$}}))$
is the Fermi-Dirac distribution function, $d$ is the dimensionality
of the system,
and we are assuming an isotropic parabolic band system.
Clearly, for the case when that temperature $T$ is small on the scale of
the energy scale over which $\tau(\varepsilon)$ varies significantly,
$\langle \tau \rangle \approx [\langle\tau^{-1}\rangle]^{-1}$.
Conversely, in the case where $T$ is large on the scale over which
$\tau(\varepsilon)$ varies, there can be significant differences
in the calculated mobility using the two different methods.
For example, we show below that in particular cases in GaAs
quantum wells, $\mu_0/\mu_\infty \approx 2$.  Thus, for accurate
theoretical determination of the mobility at the semiclassical level,
it is important that $e-e$ scattering effects are included.
Furthermore, it has been shown\cite{flen95b} that
experiments measuring the drag rate between electron gases between
two coupled quantum wells is sensitive of the exact details of the
linear response distribution function $f$ in each layer.
Since $f$ is strongly affected by $e-e$ scattering in the intermediate
temperature regime $T\approx 0.5\, T_F$ (where $T_F$ is the Fermi
temperature), it is important to include the effects of $e-e$ scattering
in calculations of the drag rate.

In this paper, we demonstrate an efficient way of including $e-e$
scattering in the calculation of for linear transport for
two-dimensional cylindrically symmetric systems,
within the semiclassical Boltzmann equation formalism.
A similar calculation has been
presented in three-dimensions.\cite{sanb95}  Within this formalism,
the Boltzmann equation for linear response can be solved exactly
(within numerical accuracy).
We have included full effects of finite-temperature dynamical screening,
which automatically includes phenomena such as Landau damping and
collective mode enhancements to scattering.
The description of the $e-e$ scattering formalism for
the Boltzmann equation is given in sections \ref{sec:formal}
and \ref{sec:scatop},
and section \ref{sec:results} contains the results and discussion.
Throughout this paper, we assume that bands are isotropic
and parabolic.

%%%%%%%%%%%%%%%% FORMALISMA %%%%%%%%%%%%%%%%%%%%%%%%%%%%%%%%%%%%%

\section{Electron--electron scattering probability}
\label{sec:formal}

The $e-e$ scattering occurs in the presence of other conduction
electrons, and hence the bare interparticle Coulomb interaction
$U(q)$ is screened.
Furthermore, at the intermediate temperatures in which we are interested,
the energy transfer between the electrons in a scattering event is
often a substantial fraction of the kinetic energy of the electrons,
and hence the scattering matrix elements for $e-e$ interactions
should be calculated using the {\em dynamically} screened Coulomb
interaction $V(q,\omega) = U(q)/\epsilon(q,\omega)$, where
$\epsilon(q,\omega)$ is the dielectric function.
In this paper, we use $\epsilon(q,\omega)$ given by
the random phase approximation (RPA), which we evaluate using
a method described previously by us,\cite{flen95b} and we use the
Born approximation for the scattering probability.

The scattering probability $w(\mbox{\boldmath{$k$}}_1' \sigma_1,
\mbox{\boldmath{$k$}}_2' \sigma_2;
\mbox{\boldmath{$k$}}_1 \sigma_1, \mbox{\boldmath{$k$}}_2 \sigma_2)$ for a pair
of electrons initially
in states $\mbox{\boldmath{$k$}}_1\sigma_1,\mbox{\boldmath{$k$}}_2\sigma_2$ to
be scattered to
$\mbox{\boldmath{$k$}}_1'\sigma_1,\mbox{\boldmath{$k$}}_2'\sigma_2$ depends on
whether
or not the electrons have the same or opposite spins.
For electrons with the same spin, say $\uparrow$,
\begin{equation}
w(\mbox{\boldmath{$k$}}+\mbox{\boldmath{$q$}}\uparrow,
\mbox{\boldmath{$k$}}'-\mbox{\boldmath{$q$}}\uparrow;
\mbox{\boldmath{$k$}}\uparrow,\mbox{\boldmath{$k$}}'\uparrow)
= \frac{1}{2}\ \frac{2\pi}{\hbar}
|V(\mbox{\boldmath{$q$}},\varepsilon_{\mbox{\boldmath{$k$}}
+\mbox{\boldmath{$q$}}}-\varepsilon_{\mbox{\boldmath{$k$}}}) - V
(\mbox{\boldmath{$k$}}'-\mbox{\boldmath{$q$}}+\mbox{\boldmath{$k$}},
\varepsilon_{\mbox{\boldmath{$k$}}'-\mbox{\boldmath{$q$}}}
-\varepsilon_{\mbox{\boldmath{$k$}}})|^2,
\label{likespin}.
\end{equation}
The fraction $1/2$ in Eq. (\ref{likespin}) is due to double counting, since
$w(\mbox{\boldmath{$k$}}+\mbox{\boldmath{$q$}}\uparrow,
\mbox{\boldmath{$k$}}'-\mbox{\boldmath{$q$}}\uparrow;
\mbox{\boldmath{$k$}}\uparrow,\mbox{\boldmath{$k$}}'\uparrow)$ and
$w(\mbox{\boldmath{$k$}}'-\mbox{\boldmath{$q$}}\uparrow,
\mbox{\boldmath{$k$}}+\mbox{\boldmath{$q$}}\uparrow;
\mbox{\boldmath{$k$}}\uparrow,\mbox{\boldmath{$k$}}'\uparrow)$
describe exactly the same process.
For opposite spins,
\begin{equation}
w(\mbox{\boldmath{$k$}}+\mbox{\boldmath{$q$}}\uparrow,
\mbox{\boldmath{$k$}}'-\mbox{\boldmath{$q$}}\downarrow;
\mbox{\boldmath{$k$}}\uparrow,\mbox{\boldmath{$k$}}'\downarrow)
= \frac{2\pi}{\hbar}
|V(\mbox{\boldmath{$q$}},\varepsilon_{\mbox{\boldmath{$k$}}
+\mbox{\boldmath{$q$}}}-\varepsilon_{\mbox{\boldmath{$k$}}})|^2.
\label{unlikespin}
\end{equation}
There is an equal probability that an electron scatters off
another electron with equal or opposite spin, so
one can sum over the Eqs. (\ref{likespin}) and (\ref{unlikespin})
to obtain an ``average" scattering probability\cite{exchange}
\begin{eqnarray}
\overline{w}(\mbox{\boldmath{$k$}}+\mbox{\boldmath{$q$}},
\mbox{\boldmath{$k$}}'-\mbox{\boldmath{$q$}};\mbox{\boldmath{$k$}},
\mbox{\boldmath{$k$}}')
&=& \frac{2\pi}{\hbar}\Bigl\{
|V(\mbox{\boldmath{$q$}},\varepsilon_{\mbox{\boldmath{$k$}}
+\mbox{\boldmath{$q$}}}-\varepsilon_{\mbox{\boldmath{$k$}}})|^2 \nonumber\\
&&\phantom{\ \ \ }-
\frac{1}{2}\mbox{Re}[V(\mbox{\boldmath{$q$}},
\varepsilon_{\mbox{\boldmath{$k$}}+\mbox{\boldmath{$q$}}}-
\varepsilon_{\mbox{\boldmath{$k$}}})\,
V^*(\mbox{\boldmath{$k$}}'-\mbox{\boldmath{$q$}}-\mbox{\boldmath{$k$}},
\varepsilon_{\mbox{\boldmath{$k$}}'-\mbox{\boldmath{$q$}}}
-\varepsilon_{\mbox{\boldmath{$k$}}})]\Bigr\}.
\label{averagew}
\end{eqnarray}
The first and second terms in Eq.\ (\ref{averagew}) referred to as
the direct and exchange terms, respecitively.

In practice, the exchange term often makes calculations considerably
more complicated and is usually ignored.
The physical grounds for doing so are as
follows.
First, $e-e$ collisions are usually dominated by small $q$ scattering
(because $V(q,\varepsilon)$ falls quickly with $q$ for finite
$\varepsilon$).  The direct
[exchange] term has the form $|V(q,\varepsilon)|^2$
[$V(q,\varepsilon) V(q',\varepsilon')$ with $q'\gg q$], which implies
that the direct term usually dominates over the exchange.  Also,
the sign of exchange term can sometimes be negative,
which leads to cancellation of this term within collision integral.
The effect of the exchange term was studied in a 3D system with
scatically screened interaction.\cite{coll93} There it was found that
the exchange term was significant for
$na_0^3=1$ but not for $na_0^3=0.1$.
In the calculations that follows we have used $na_0^2= 0.15$,
and since we furthermore include dynamical screening,
which leads to a peaked interaction
at small $q$, we can assume that the direct interaction
dominates in our case.

We write down the formal expressions for the electron--electron
scattering operator both including and excluding the exchange
term.  However, in the actual numerical evaluation, we ignore the
exchange interaction.

\section{Electron--electron scattering operator}
\label{sec:scatop}

The Boltzmann equation for electrons in uniform electric field
$\mbox{\boldmath{$E$}}$
producing a force $\mbox{\boldmath{$F$}} = (-e)\mbox{\boldmath{$E$}}$ is
\begin{equation}
\hbar^{-1}\mbox{\boldmath{$F$}}\cdot\frac{\partial f}{\partial
\mbox{\boldmath{$k$}}} =
\left(\frac{\partial f}{\partial t}\right)_{e-e} +
\left(\frac{\partial f}{\partial t}\right)_{\mathrm{p,i}}
\end{equation}
where the subscripts ${e-e}$ and p,i are for
scattering due to electron--electron interactions and
the phonon+impurity interactions, respectively.

We define the function $\Psi(\mbox{\boldmath{$k$}})$, related to the deviation
of the distribution function from equilibrium, as
\begin{equation}
f(\mbox{\boldmath{$k$}})-f_0(k) \equiv f_0(k)[1-f_0(k)]
\Psi(\mbox{\boldmath{$k$}}).
\end{equation}
This function can be written in terms of a sum of angular components
\begin{equation}
\Psi(k,\theta) = \sum_n \psi_n(k)\ \cos(n\theta).
\end{equation}
where $\theta$ is the angle from an axis of symmetry (here, the
direction of the electric field).
By the assumption of cylindrical symmetry of the system,
the scattering terms in the Boltzmann equation do not mix
different $\cos(n\theta)$ components,
and one can isolate and concentrate on the
$\cos\theta$ component, $\psi_1(k)$,
since this is the one that affects the current and hence the mobility.

The $\cos\theta$ component of the linearized
electron--electron collision operator (i.e., neglecting higher powers
in $\Psi(\mbox{\boldmath{$k$}})$), which we denote
$I_{e-e}[\psi_1]$, is\cite{smith}
\begin{eqnarray}
I_{e-e}[\psi_1 x_{\mbox{\boldmath{$k$}},\mbox{\boldmath{$F$}}}](k) &=&
-2\int \frac{d\mbox{\boldmath{$k$}}'}{(2\pi)^2} \int
\frac{d\mbox{\boldmath{$q$}}}{(2\pi)^2}\
\overline{w}(\mbox{\boldmath{$k$}}+\mbox{\boldmath{$q$}},
\mbox{\boldmath{$k$}}'-\mbox{\boldmath{$q$}};\mbox{\boldmath{$k$}},\mbox{\boldmath{$k$}}')\times \nonumber\\
&&\phantom{xxx}  f^0(\mbox{\boldmath{$k$}}) f^0(\mbox{\boldmath{$k$}}')
\Bigl[1 - f^0(\mbox{\boldmath{$k$}}+\mbox{\boldmath{$q$}}) \Bigr] \Bigl[1 -
f^0(\mbox{\boldmath{$k$}}'-\mbox{\boldmath{$q$}})\Bigl]\nonumber\\
&&\phantom{xxx} \delta\left(\epsilon_{\mbox{\boldmath{$k$}}} +
\epsilon_{\mbox{\boldmath{$k$}}'} -
\epsilon_{\mbox{\boldmath{$k$}}+\mbox{\boldmath{$q$}}} -
\epsilon_{\mbox{\boldmath{$k$}}' - \mbox{\boldmath{$q$}}}\right)  \nonumber\\
&&\phantom{xxx} \Bigl\{\psi_1(k)
x_{\mbox{\boldmath{$k$}},\mbox{\boldmath{$F$}}}  +
\psi_1(k')x_{\mbox{\boldmath{$k$}}',\mbox{\boldmath{$F$}}} \nonumber\\
&&\phantom{xxx} -
\psi_1(|\mbox{\boldmath{$k$}}+\mbox{\boldmath{$q$}}|)
x_{\mbox{\boldmath{$k$}}+\mbox{\boldmath{$k$}},\mbox{\boldmath{$F$}}}
-
\psi_1(|\mbox{\boldmath{$k$}}'-\mbox{\boldmath{$q$}}|)
x_{\mbox{\boldmath{$k$}}'-\mbox{\boldmath{$q$}},\mbox{\boldmath{$F$}}}
\Bigr\}.
\label{eecoll}
\end{eqnarray}
Here, $x_{\mbox{\boldmath{$k$}},\mbox{\boldmath{$k$}}'}$ is the cosine of the
angle between
$\mbox{\boldmath{$k$}}$ and $\mbox{\boldmath{$k$}}'$.
The goal is to write the operator $I$ in the form
\begin{equation}
I[\psi_1 x_{\mbox{\boldmath{$k$}},\mbox{\boldmath{$F$}}}](k) =
x_{\mbox{\boldmath{$k$}},\mbox{\boldmath{$F$}}}\int_0^\infty dp\ \ p\; K(k,p)\
\psi_1(p).
\end{equation}
The kernal $K(k,p)$ is symmetric, from detailed balance,\cite{smith}
and the extra factor of $p$ in the integral comes from phase-space.
Thus, in order to incorporate electron--electron scattering
for a particular density and temperature into a calculation,
one need only generate $K(k,p)$ once and store it;
it can then be used for all calculations involving
electron--electron scattering at that density and temperature.

The four $\psi$'s in Eq. (\ref{eecoll}) give four terms, each
of which give a contribution to the kernel, $K = K_1 + K_2 + K_3 +K_4$.
In the following subsections, we explicitly write down the form of
each of these kernels.

\subsection{First term, involving $\psi_1(k)$}

The $\psi_1(k) x_{\mbox{\boldmath{$k$}},\mbox{\boldmath{$F$}}}$ can be factored
out, and we obtain
\begin{eqnarray}
K_1(k,p) &=&
-\delta(k-p) k^{-1} f_0(k) \int \frac{d\mbox{\boldmath{$q$}}}{(2\pi)^2}\
[1-f_0(\mbox{\boldmath{$k$}}+\mbox{\boldmath{$q$}})]
\nonumber\\
&&\phantom{hell}2\int \frac{d\mbox{\boldmath{$k$}}'}{(2\pi)^2}
f_0(\mbox{\boldmath{$k$}}')\,[1-f_0(\mbox{\boldmath{$k$}}'-
\mbox{\boldmath{$q$}})]
\overline{w}(\mbox{\boldmath{$k$}}+\mbox{\boldmath{$q$}},
\mbox{\boldmath{$k$}}';\mbox{\boldmath{$k$}},
\mbox{\boldmath{$k$}}'+\mbox{\boldmath{$q$}})\nonumber\\
&&\phantom{hell}\delta\left(\varepsilon_{\mbox{\boldmath{$k$}}'+
\mbox{\boldmath{$q$}}} - \varepsilon_{\mbox{\boldmath{$k$}}'} -
\{\varepsilon_{\mbox{\boldmath{$k$}}+\mbox{\boldmath{$q$}}}-
\varepsilon_{\mbox{\boldmath{$k$}}}\}\right).
\end{eqnarray}

In the event where the exchange interaction can be neglected, one
obtains, as in Ref.\ \onlinecite{sanb95} (we denote the
scattering integral which neglects the exchange interaction
with an asterisk)
\begin{eqnarray}
K_1^*(k,p) &=&
-\delta(k-p) \Bigl[\frac{f^0(k)}{\pi k}
\int \frac{d\mbox{\boldmath{$q$}}}{(2\pi)^2}\;
[1-f^0(\mbox{\boldmath{$k$}}+\mbox{\boldmath{$q$}})]\;
\frac{2\pi}{\hbar}
|V(\mbox{\boldmath{$q$}},\varepsilon_{\mbox{\boldmath{$k$}}+
\mbox{\boldmath{$q$}}}-\varepsilon_{\mbox{\boldmath{$k$}}})|^2\nonumber\\
&&\ \ \ \
\mathrm{Im}[\chi(\mbox{\boldmath{$q$}},\varepsilon_{\mbox{\boldmath{$k$}}} -
\varepsilon_{\mbox{\boldmath{$k$}}+\mbox{\boldmath{$q$}}})]
\;
n_B(\varepsilon_{\mbox{\boldmath{$k$}}+\mbox{\boldmath{$q$}}}
-\varepsilon_{\mbox{\boldmath{$k$}}})\Bigr].
\end{eqnarray}
where
$n_B(\varepsilon) = [\exp(\beta\varepsilon) - 1]^{-1}$ is the Bose
function, and $\chi(q,\omega)$ is the RPA polarizability.

\subsection{Second term, involving $\psi_1(k')$}

Since $\cos(\theta+\theta') = \cos(\theta)\cos(\theta') -
\sin(\theta)\sin(\theta')$ and the $\sin$ terms vanish from symmetry
considerations, we can write $x_{\mbox{\boldmath{$k$}}',\mbox{\boldmath{$F$}}}
=  x_{\mbox{\boldmath{$k$}},\mbox{\boldmath{$F$}}}
x_{\mbox{\boldmath{$k$}}',\mbox{\boldmath{$k$}}}$.
Then, the second kernel is
\begin{eqnarray}
K_2(k,p)
&=& - 2 \frac{f^0(k)f_0(p) }{(2\pi)^2}
\int_0^{2\pi} d\theta_{\mbox{\boldmath{$k$}},\mbox{\boldmath{$p$}}}\
\cos\theta_{\mbox{\boldmath{$k$}},\mbox{\boldmath{$p$}}}
\nonumber\\
&&\phantom{\ \ } \int \frac{d\mbox{\boldmath{$q$}}}{(2\pi)^2}
w(\mbox{\boldmath{$k$}}+\mbox{\boldmath{$q$}},\mbox{\boldmath{$p$}}
-\mbox{\boldmath{$q$}};\mbox{\boldmath{$k$}},\mbox{\boldmath{$p$}})
[1-f^0(\mbox{\boldmath{$k$}}+\mbox{\boldmath{$q$}})]
[1-f^0(\mbox{\boldmath{$p$}}-\mbox{\boldmath{$q$}})]\nonumber\\
&&\phantom{\ \ \ \ \ }\delta(\varepsilon_{\mbox{\boldmath{$k$}}} +
\varepsilon_{\mbox{\boldmath{$p$}}} -
\varepsilon_{\mbox{\boldmath{$k$}}+\mbox{\boldmath{$q$}}} -
\varepsilon_{\mbox{\boldmath{$k$}}-\mbox{\boldmath{$q$}}}).
\label{i2}
\end{eqnarray}

The $\mbox{\boldmath{$q$}}$ integration can be evaluated by the change of
variables
\begin{eqnarray}
\mbox{\boldmath{$q$}} &=& \mbox{\boldmath{$Q$}} +
\Delta\mbox{\boldmath{$k$}}/2,\nonumber\\
\Delta\mbox{\boldmath{$k$}} &=&
\mbox{\boldmath{$p$}}-\mbox{\boldmath{$k$}},\nonumber\\
\overline{\mbox{\boldmath{$k$}}} &=&
(\mbox{\boldmath{$k$}}+\mbox{\boldmath{$p$}})/2.
\end{eqnarray}
Then, the $\delta$-function in Eq.\ (\ref{i2}) becomes
\begin{equation}
\delta(\varepsilon_{\mbox{\boldmath{$k$}}}+\varepsilon_{\mbox{\boldmath{$p$}}}
- \varepsilon_{\mbox{\boldmath{$k$}}+\mbox{\boldmath{$q$}}}
-\varepsilon_{\mbox{\boldmath{$p$}}-\mbox{\boldmath{$q$}}}) =
\delta\left(\frac{\hbar^2}{m}
\left[Q^2 - \left(\frac{\Delta\mbox{\boldmath{$k$}}}{2}\right)^2\right]\right),
\end{equation}
which gives
\begin{eqnarray}
K_2(k,p)
&=& -  \frac{m f^0(k)f_0(p)}{8\pi^4\hbar^2}
\int_0^{\pi} d\theta_{\mbox{\boldmath{$k$}},\mbox{\boldmath{$p$}}}\
\cos\theta_{\mbox{\boldmath{$k$}},\mbox{\boldmath{$p$}}}
\nonumber\\
&&\phantom{\ \ }
\int_0^{2\pi} d\phi\
w(\overline{\mbox{\boldmath{$k$}}}+\mbox{\boldmath{$Q$}},
\overline{\mbox{\boldmath{$k$}}}-\mbox{\boldmath{$Q$}};
\overline{\mbox{\boldmath{$k$}}}+
\Delta\mbox{\boldmath{$k$}},\overline{\mbox{\boldmath{$k$}}}
-\Delta\mbox{\boldmath{$k$}})\nonumber\\
&&\phantom{\ \ \ \ }
[1-f^0(\varepsilon_{\overline{\mbox{\boldmath{$k$}}}+\mbox{\boldmath{$Q$}}})]\;
[1-f^0(\varepsilon_{\overline{\mbox{\boldmath{$k$}}}-\mbox{\boldmath{$Q$}}})],
\end{eqnarray}
where $\phi$ is the angle between $\mbox{\boldmath{$Q$}}$ and
$\overline{\mbox{\boldmath{$k$}}}$.

\subsection{Third term, involving
$\psi_1(|\mbox{\boldmath{$k$}}+\mbox{\boldmath{$q$}}|)$}

Using $x_{\mbox{\boldmath{$k$}}+\mbox{\boldmath{$q$}},\mbox{\boldmath{$F$}}} =
x_{\mbox{\boldmath{$k$}},\mbox{\boldmath{$F$}}}
x_{\mbox{\boldmath{$k$}},\mbox{\boldmath{$k$}}+\mbox{\boldmath{$q$}}}$ in the
integrand
and letting $\mbox{\boldmath{$p$}} =
\mbox{\boldmath{$k$}}+\mbox{\boldmath{$q$}}$, gives
\begin{eqnarray}
K_3(k,p) &=& \frac{2f^0(k)[1-f^0(p)]}{(2\pi)^2} \int_0^{2\pi}
d\theta_{\mbox{\boldmath{$p$}},\mbox{\boldmath{$k$}}}\
\cos\theta_{\mbox{\boldmath{$p$}},\mbox{\boldmath{$k$}}}
\phantom{\ \ \ }\int \frac{d\mbox{\boldmath{$k$}}'}{(2\pi)^2}
f^0(\mbox{\boldmath{$k$}}'+\mbox{\boldmath{$p$}}-\mbox{\boldmath{$k$}})
[1-f^0(\mbox{\boldmath{$k$}}')]
\nonumber\\
&&\overline{w}(\mbox{\boldmath{$p$}},\mbox{\boldmath{$k$}}';
\mbox{\boldmath{$k$}},\mbox{\boldmath{$k$}}'+\mbox{\boldmath{$p$}}
-\mbox{\boldmath{$k$}})
\delta(\varepsilon_{\mbox{\boldmath{$p$}}} +
\varepsilon_{\mbox{\boldmath{$k$}}'} - \varepsilon_{\mbox{\boldmath{$k$}}}
-\varepsilon_{\mbox{\boldmath{$k$}}'+\mbox{\boldmath{$p$}}-
\mbox{\boldmath{$k$}}}).
\end{eqnarray}
The $\delta$-function $\delta(\hbar^2
(\mbox{\boldmath{$p$}}-\mbox{\boldmath{$k$}})\cdot
(\mbox{\boldmath{$k$}}'-\mbox{\boldmath{$k$}})/m)$
reduces the $d\mbox{\boldmath{$k$}}'$ dimensional integral to one dimension.

If one neglects exchange, then as with the first term the
$d\mbox{\boldmath{$k$}}'$
integral can be done, giving
\begin{eqnarray}
K_3^*(k,p) &=& \frac{1}{2\pi^3}
\frac{[f^0(\varepsilon_{\mbox{\boldmath{$p$}}})-
f^0(\varepsilon_{\mbox{\boldmath{$k$}}})]}
{4 \sinh^2[(\varepsilon_{\mbox{\boldmath{$p$}}}-
\varepsilon_{\mbox{\boldmath{$k$}}})/(2k_B T)]}
\nonumber\\
&&\int_0^\pi d\theta_{\mbox{\boldmath{$k$}},\mbox{\boldmath{$p$}}}
cos\theta_{\mbox{\boldmath{$k$}},\mbox{\boldmath{$p$}}}
\
\frac{2\pi}{\hbar}|V(\mbox{\boldmath{$p$}}-
\mbox{\boldmath{$k$}},\varepsilon_{\mbox{\boldmath{$k$}}}-
\varepsilon_{\mbox{\boldmath{$p$}}})|^2\;
\mbox{Im}[\chi(\mbox{\boldmath{$p$}}-\mbox{\boldmath{$k$}},
\varepsilon_{\mbox{\boldmath{$k$}}}-\varepsilon_{\mbox{\boldmath{$p$}}})]
\end{eqnarray}

\subsection{Fourth term, involving
$\psi_1(|\mbox{\boldmath{$k$}}'-\mbox{\boldmath{$q$}}|)$}

The kernel is
\begin{eqnarray}
K_4(k,p) &=& \frac{2f^0(k)}{(2\pi)^2} \int_0^{2\pi}
d\theta_{\mbox{\boldmath{$p$}},\mbox{\boldmath{$k$}}}
\cos(\theta_{\mbox{\boldmath{$p$}},\mbox{\boldmath{$k$}}}) \nonumber\\
&&\phantom{ \ \ }\left\{\int \frac{d\mbox{\boldmath{$q$}}}{(2\pi)^2}\
w(\mbox{\boldmath{$k$}}+\mbox{\boldmath{$q$}},\mbox{\boldmath{$k$}}'
-\mbox{\boldmath{$q$}};\mbox{\boldmath{$k$}},\mbox{\boldmath{$k$}}')
[1-f^0(\mbox{\boldmath{$k$}}+\mbox{\boldmath{$q$}})]\
f^0(\mbox{\boldmath{$p$}}+\mbox{\boldmath{$q$}})\right.\nonumber\\
&&\phantom{\ \ \ }\left.\delta(\varepsilon_{\mbox{\boldmath{$k$}}} +
\varepsilon_{\mbox{\boldmath{$p$}}+\mbox{\boldmath{$q$}}} -
\varepsilon_{\mbox{\boldmath{$k$}} + \mbox{\boldmath{$q$}}} -
\varepsilon_{\mbox{\boldmath{$p$}}})\right\}
\end{eqnarray}
The term in the $\delta$-function goes as
\begin{equation}
\varepsilon_{\mbox{\boldmath{$p$}}+\mbox{\boldmath{$q$}}} -
\varepsilon_{\mbox{\boldmath{$p$}}} + \varepsilon_{\mbox{\boldmath{$k$}}}
-\varepsilon_{\mbox{\boldmath{$k$}}+\mbox{\boldmath{$q$}}} =
\frac{\hbar^2}{m}\left\{ \mbox{\boldmath{$q$}}\cdot
(\mbox{\boldmath{$p$}}-\mbox{\boldmath{$k$}})\right\},
\end{equation}
which reduces the $\mbox{\boldmath{$q$}}$ integration down to one dimension.

In fact, the kernels for higher order components are very similar
to the ones given above.  For an angular variation proportional to
$\cos n\theta$, the $K_1$ term is identical for all $n$, whereas
with $K_2$, $K_3$ and $K_4$ one simply replaces $\cos\theta$
with $\cos n \theta$ in the $\theta$ integration.

We have shown that one can calculate the matrix $K(k,p)$ which
gives the electron--electron scattering term for small deviations
from the equilibrium.  Once the matrix $K(k,p)$ has been calculated,
one simply needs to iterate the equation for $\psi_1(k)$ until
convergence is obtained.

In order to calculate $\psi_1(k)$ for the case when the $e-e$ scattering
rate dominates, it is often useful to use the fact that
electron--electron scattering leaves a drifted Fermi-Dirac distribution
invariant.\cite{test}
Thus, for the case of elastic or quasi-elastic collisions,
one can define
\begin{equation}
\tilde{\psi}_1(k) = \psi_1(k) - \psi_1^{\mathrm{DF}}(k)
\end{equation}
where $\psi^{\mathrm{DF}}_1(k) = \hbar k v_d /(k_B T)$ is for a drifted
Fermi-Dirac distribution. Any $v_d$ can be used.
In our case, because we cut off the matrix $K(k,p)$ at a
$k_{\mathrm{max}}$, which implicitly sets $\tilde{\psi}_1(k >
k_{\mathrm{max}}) = 0$, we chose $v_d$ so that $\tilde{\psi}_1(
k_{\mathrm{max}}) = 0$ so that the distribution function is
continuous at $k_{\mathrm{max}}$.

We write the linearized $e-e$ scattering term as
\begin{equation}
I_{e-e}[\psi_1(k)] =
-\frac{\psi(k)}{\tau_{ee}} + J[\psi_1(k)]
\end{equation}
where the first term on the right hand side corresponds to the
diagonal $K_1$ term and $J$ corresponds to $K_2 +K_3 + K_4$.
The Boltzmann equation for $\psi_1(k)$ in the case when
the other scattering mechanisms are quasielastic
(which might include acoustic phonon scattering, which generally involves
very small energy electron loss) is
\begin{equation}
eEv(k)\left(\frac{\partial f_0}{\partial\varepsilon}\right) =
-\frac{f_0(k)[1-f_0(k)]\left(\tilde{\psi}_1(k)+
\psi^{\mathrm{DF}}_1(k)\right)}
{\tau_{\mathrm{el}}(k)}
-\frac{\tilde{\psi}_1(k)}{\tau_{e-e}(k)} + J[\tilde{\psi}_1](k),
\end{equation}
where $\tau_{\mathrm{el}}^{-1}(k)$ is the quasi-elastic scattering rate.
This implies that one must iterate the equation
\begin{equation}
\tilde{\psi}_1(k)
= \frac{
\frac{eEv(k)}{k_B T} f_0(k)[1-f_0(k)] + I^*[\tilde{\psi}_1] -
f_0(k)[1-f_0(k)] \psi^{\mathrm{DF}}_1(k)\tau^{-1}_{\mathrm{el}}(k)}
{f_0(k)[1-f_0(k)]\tau^{-1}_{\mathrm{el}}(k) + \tau_{e-e}^{-1}(k) }
\end{equation}
to find $\psi_1(k)$.

\section{Results and discussion}
\label{sec:results}

We study the case of electrons confined in a 100 \AA\ wide
square GaAs quantum well with infinite barriers.
We assume that the there is a $\delta$-doping
layer of (uncorrelated) charged impurities, equal in density to that of
the electrons in the well, situated a distance $d$ away from
the center of the well.  We included three scattering mechanisms:
$e-e$, charged-impurity and acousitic-phonon scattering, and we approximated
the acousitic-phonon scattering as being elastic.

We calculated the matrix in the form on 200 by 200 grid-points from
$k = 0$ to $k = 5k_F$, and we used spline routines to interpolate
between the grid points.  The $K_3(k,p)$ and $K_4(k,p)$
diverge logarithmically as $k\rightarrow p$, which complicates
the splining procedure, but we got around this problem by splining
$K_{3,4}(k,p)/\ln(|k-p|)$, which is a smooth function.

In Fig. 1, we show the deviation function
$\psi_1(k)/\sqrt{k}$ for a fixed density $n=1.5\times
10^{11}\;\mbox{cm}^{-2}$ and temperature $T=30\;\mbox{K}$,
but with several different distances $d$ of the ionized impurities from the
center of the quantum well.  Note that when the distribution is
a drifted Fermi-Dirac function, $\psi_1(k)/\sqrt{k} =
\mbox{constant}$.  Thus, as the impurities are moved further
away, the impurity scattering becomes weaker and
the $e-e$ scattering starts to dominate\cite{weak}
and drive the distribution function closer to a drifted Fermi-Dirac function.
The inset shows $\psi_1(k)$ calculated both including and
excluding electron--electron scattering for $d=100\;$\AA, which
shows more clearly the effect of $e-e$ on $\psi_1(k)$.
While transport experiments in a single layer are not particularly
dependent on the details of the shape of $\psi_1(k)$,
it has been shown\cite{flen95b} that drag experiments
in coupled quantum wells are quite sensitive to the details of
$\psi_1(k)$.  In particular, when $\psi_1(k)$
rises faster than $\sqrt k$
(which implies that there are more carriers in the high energy region
than for a drifted Fermi-Dirac distribution), the drag rate increases
because high energy particles give a larger contribution to the
overall drag rate, and there is greater opportunity for coupling
to the plasmons of the system, which also enhances the drag rate.
Therefore, for the purpose of calculating the drag rate in coupled quantum
wells in intermediate temperatures, it is crucial to calculate the
actual form of $\psi_1(k)$ accurately, including all salient
scattering mechanisms.

Fig. 2 shows the mobility $\mu$ as a function of ionized impurity
distance $d$ from the center of the quantum well.  Also shown are the
mobilities $\mu_0 = e\langle \tau\rangle/m$ and
$\mu_\infty = e/(\langle\tau^{-1}\rangle m)$,
for the limits of zero and infinite
electron--electron scattering, respectively.
The $\mu_0$ is generally larger than $\mu_\infty$ because $e-e$
scattering tends to scatter ``runaway" electrons with large velocities
(where the impurity scattering rate is small) back into lower
velocity states.
The inset shows that $\mu_0$ for this case can be almost twice
$\mu_\infty$.

As $d$ becomes larger, the $e-e$ scattering dominates over all
other scattering mechanisms and $\mu \rightarrow \mu_\infty$.
Conversely, for small $d$, the impurity scattering is relatively
large compared to the electron scattering, and $\mu$ is
closer to the $\mu_0$ than $\mu_\infty$.  The crossover from
$\mu_0$ to $\mu_\infty$ is shown with the open squares
in the inset of Fig. 2.  The crossover occurs
when the impurity scattering and electron--electron mean
free paths become equivalent.
The transport scattering rate,
for impurity scattering is given by $\tau_{tr}^{-1} =
e/(\mu m) = 2.6 \times 10^{16}\;\mbox{cm$^2$V$^{-1}$s$^{-2}$}/\mu$.
The electron--electron scattering in two-dimensional systems
is approximately given by\cite{giul82,strictly}
\begin{equation}
\tau_{e-e}^{-1} \approx \frac{E_F}{\hbar} \left[\frac{k_B T}
{E_F}\right]^2 \left[\ln\left[\frac{E_F}{k_B T}\right] + \ln
\left[2\frac{q_{TF}}{p_F}\right] + 1 \right].
\end{equation}
For this system, this is on the order of $10^{12}\,\mbox{s}^{-1}$.
Thus, the crossover point, which should occur when these
two are equal, is given by $\mu \approx 3\times 10^4 \mbox{cm}^2
\mbox{V}^{-1}\mbox{s}^{-1}$.  An inspection of Fig. 2 also
shows this to hold.
Chabasseur-Molyneux {\it et al.}\cite{chab95} have experimentally
found this crossover in GaAs/AlGaAs heterojunctions.
Finally, at $d = 150\;$\AA, $\mu \approx 0.6\mu_0$, implying
the $e-e$ scattering has caused a substantial reduction in the mobility.

To summarize, in this paper we have described a method
of including electron--electron scattering, including full
finite-temperature dynamical screening, exactly in the Boltzmann equation,
for small deviations of the distribution function from equilibrium.
Using this method to calculate the distribution function and
mobilities for electrons in a GaAs quantum well, we find a
well-defined crossover from $\mu_0$ to $\mu_\infty$ (which can be
significantly different from each other) when the $e-e$
and impurity scattering mean free paths are equivalent.
For certain parameters studied, $e-e$ is responsible for reduction
in the mobility of up to 40\%.

\section{Acknowledgements}

We thank Karim El-Sayed for useful discussions.
KF was supported by the Carlsberg Foundation.

%%%%%%%%%%%%%%%% FIGURES %%%%%%%%%%%%%%%%%%%%%%%%%%%%%%%%%%%

\begin{figure}
\caption{
Deviation function $\psi_1(k) k^{-1/2}$ (normalized to 1 at $k=0$)
for electrons responding to a weak static electric field in a
GaAs quantum well, width 100\AA,
density $n=1.5\times 10^{11}\,\mbox{cm}^{-2}$ and
temperature $T=30\,\mbox{K}$, for distances $d=50, 150, 250\
\mbox{and}\ 350$ \AA\ of the charged impurity layer from
the center of the well.
The collision term includes screened impurity,
acoustic phonon and electron--electron scattering terms.
The further the impurities are moved away from
well, the more dominant the electron--electron scattering becomes
and the distribution tends to a drifted Fermi-Dirac (a pure drifted
Fermi-Dirac is a straight horizontal line).
The solid (dashed) curve in the inset shows the
deviation function (in arbitrary units)
for $d = 100$ \AA\ calculated including (excluding)
electron--electron scattering.
}
\label{fig:psi}
\end{figure}

\begin{figure}
\caption{
Mobility as a function of impurity distance from the center
of the well.  Parameters as in Fig. \protect\ref{fig:psi}.
Crosses are for $\mu$ calculated from the Boltzmann equation,
solid line is for $\mu_0$ (no $e-e$) and dashed line is for $\mu_\infty$
(infinite $e-e$).
Inset: (i) $(\mu-\mu_{\infty})/(\mu_0-\mu_\infty)$ as a
function of impurity distance from center of well (solid line).
As the mobility passes through $\sim 3\times 10^4\;\mbox{cm}^2 V^{-1}
s^{-1}$, where $e-e$ and impurity scattering
rates are approximately comparable, the mobility crosses over from
$\mu_0$ to $\mu_\infty$.
(ii) The ratio $\mu_\infty/\mu_0$ (dashed line). This ratio
can be as small as $\approx 0.5$.
}
\label{fig:mob}
\end{figure}


\begin{thebibliography}{10}
\bibitem{test}
Seeing if a drifted Fermi-Dirac distribution gives zero in $(\partial
  f/\partial t)_{e-e}$ is in fact a useful test of one's numerical procedure.
\bibitem{weak}
The acoustic phonon scattering is much weaker than $e-e$ scattering and
  comparable with the impurity scattering only for $d\approx 800 \AA$.
\end{thebibliography}
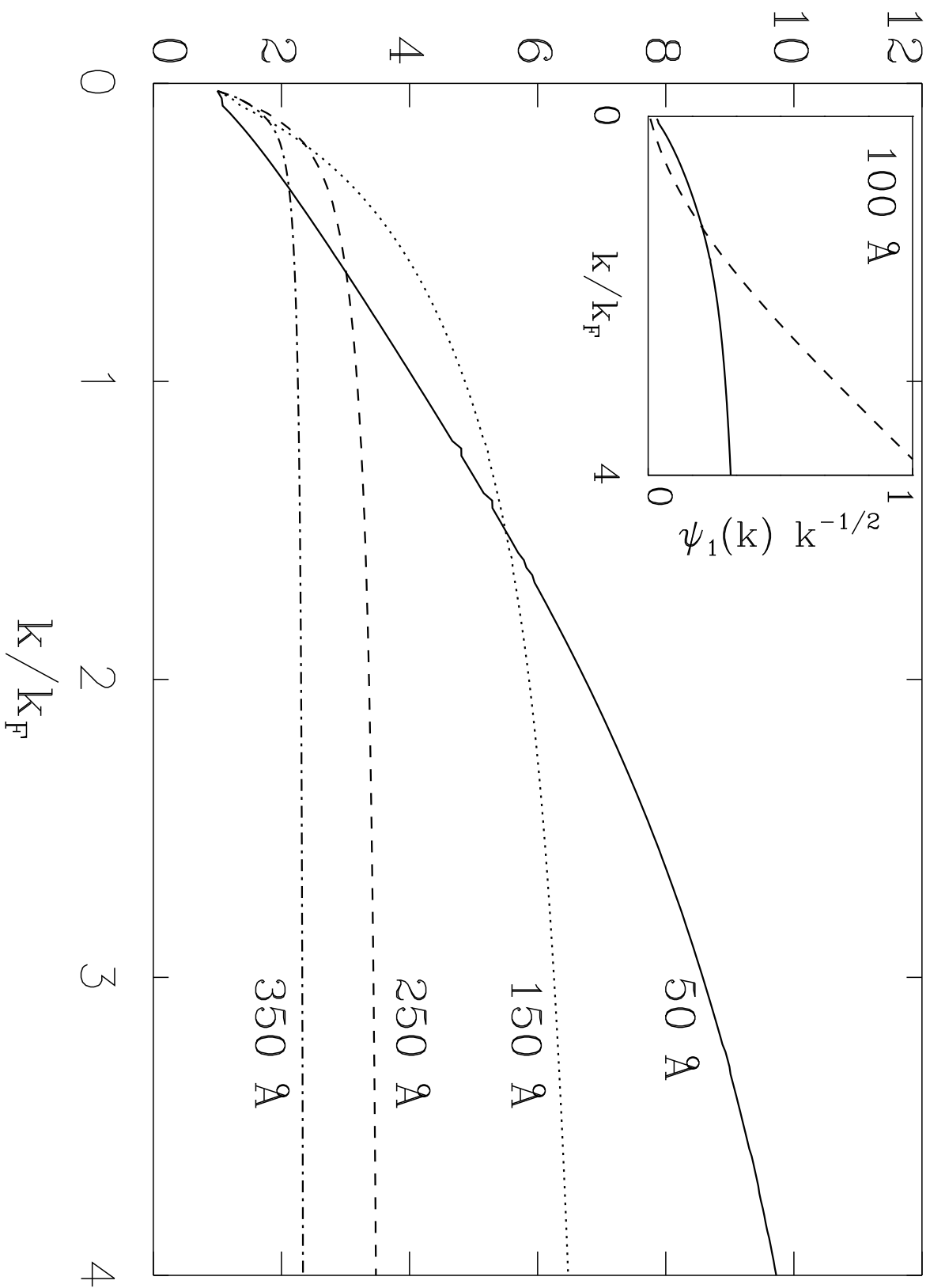
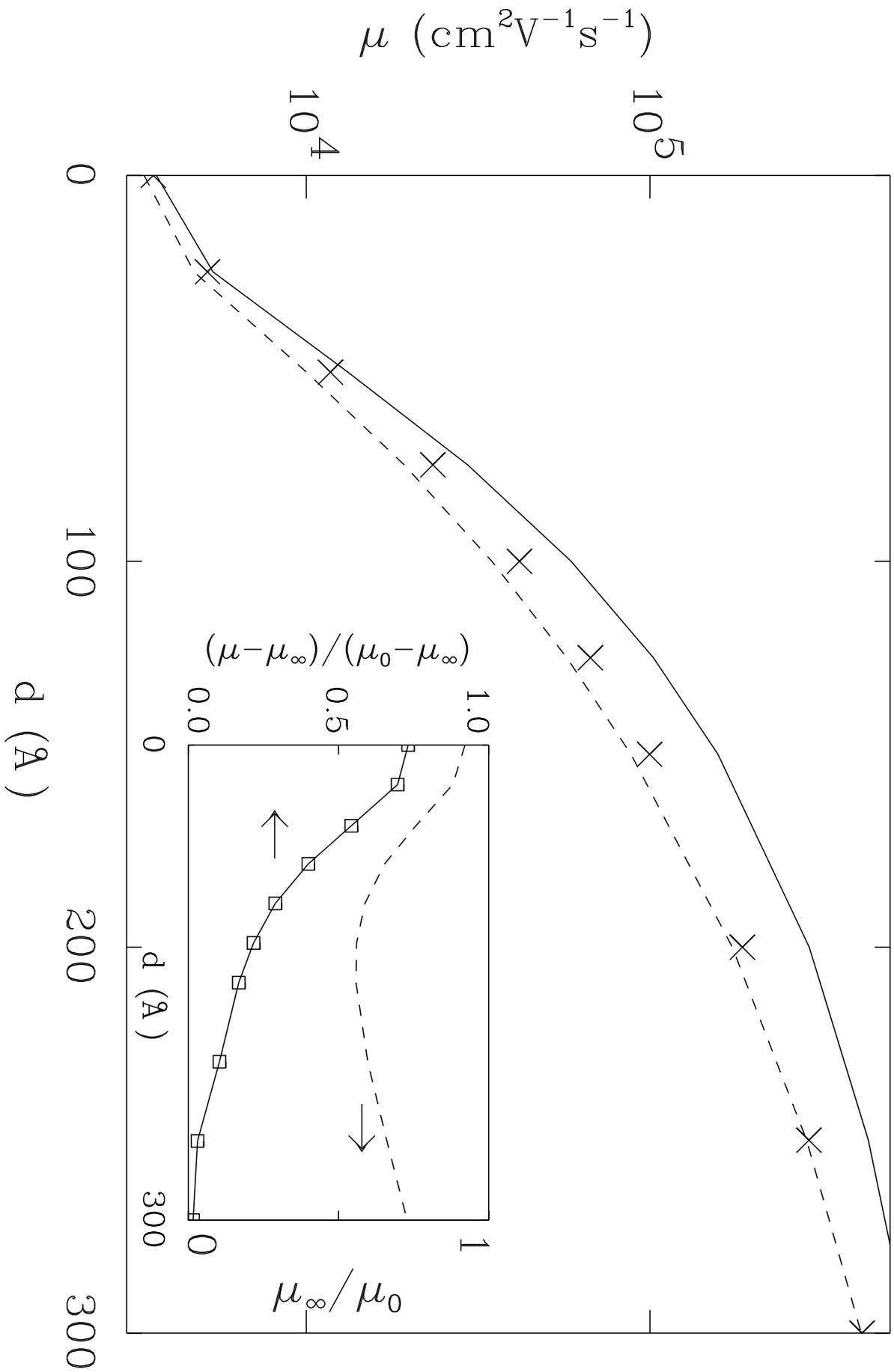
\end{document}